\documentclass[12pt,a4paper,final]{iopart}

\expandafter\let\csname equation*\endcsname\relax 
 \expandafter\let\csname endequation*\endcsname\relax 

\usepackage{amsmath}
\usepackage{iopams}  
\usepackage{graphicx}

\newcommand{\be}{\begin{equation}}
\newcommand{\ee}{\end{equation}}
\newcommand{\bea}{\begin{eqnarray}}
\newcommand{\eea}{\end{eqnarray}}
\newcommand{\non}{\nonumber}

\begin{document}

\title{Higgs Dark Energy}

\author{Massimiliano Rinaldi}
\address{Dipartimento di Fisica, Universit\`a di Trento, and TIFPA-INFN,\\  Via Sommarive 14, 38123 Povo (TN), Italy.}
\ead{massimiliano.rinaldi@unitn.it}



\begin{abstract}
\noindent We study the classical dynamics of a non-abelian Higgs theory coupled to gravity in an isotropic and homogeneous Universe.  For non-minimal coupling,  this theory leads to a model of cosmic inflation that is very attractive due to its simplicity and consistency with the latest experimental data. We show that this theory can also explain the current accelerated expansion of the Universe, provided that all the gravitational and bosonic degrees of freedom, together with their symmetries, are correctly taken in account.

 \end{abstract}



\section{Introduction}

\noindent The history of the observable Universe appears to be firmly rooted on an initial period of exponential growth, which wiped out all inhomogeneities and flattened the spacetime, known as inflation. This phase is characterized by an acceleration of the expansion rate that seems not unique in the history in the Universe. In fact, after a long period of deceleration, we know that the Universe started to accelerate again, and quite recently with respect to its age, under the action of an unknown agent called  ``dark energy''.

A non-trivial question is the nature of the acceleration mechanism. The prescriptions of general relativity are very strict when it comes to the matter content of the Universe.  There are indeed energy conditions that forbid inflation unless we assume that, during that epoch, the dynamics was driven not by ordinary matter but by a scalar field rolling over a very flat potential. This is the key idea of the first inflationary models conceived in the early 80's \cite{inflation}, which are still considered the most likely when compared to experimental data \cite{planck}. Although conceptually simple, these models have a phenomenological character: nobody knows the fundamental nature of the scalar field (called the inflaton) and its potential. 

As for dark energy, the problem is basically the same: it cannot be ordinary matter so there are several ideas trying to explain it, from modified gravity to the action of fundamental scalar fields, although the simplest explanation goes back to A.\ Einstein and consists in a very tiny as well as unnatural cosmological constant. Hopefully, future experiments like Euclid will be able to shed some light on this issue \cite{euclid}. 

The aim of this work is to explore the intriguing idea that inflation and dark energy are two sides of the same coin, namely the only fundamental scalar field known in Nature: the standard model Higgs field. That the Higgs can drive inflation is already known \cite{hinfl}. What is not known is that the same field configuration leads to a slowly varying and arbitrarily tiny effective cosmological constant in the late Universe. This conclusion emerges naturally from the analysis of the dynamics of the full model.  We remind also that there exist other non-trivial solutions with non-minimal coupling of the Higgs field to gravity, such as the Higgs monopole  with static and spherical symmetry studied in \cite{monop}.

In the standard model, the local invariance of the Higgs field is necessarily supported by non-abelian gauge fields. When gravity comes into play, the symmetries of spacetime impose restrictions on the gauge field components and one well-know example comes from cosmic strings, see e.g.\ \cite{achu}. In general, it is known that there are exact solutions for Yang-Mills theories that preserve the symmetry of spacetime, provided the gauge is fixed appropriately \cite{henneux}. Therefore, one might suspect that the gauge fields associated to the Higgs can have an important dynamical role in the expansion of the Universe, but this is not the case. In fact, there are several models where dark energy or inflation are driven by gauge fields, see for instance Refs.\ \cite{picon}-\cite{Fuzfa}. However, to produce accelerated expansion, the Yang-Mills sector  needs a non-trivial coupling to gravity and/or non-trivial potentials, none of which is the case in the standard model Higgs field, which is minimally coupled to gravity at low energy.

On the other hand, the Higgs field is a complex doublet with internal degrees freedom that respect a conserved $SU(2)$ current. The multifield dynamics of such a system was already studied  in the context of Higgs inflation, and it was shown that there are tiny departures from the simplified model of a real scalar field obtained by imposing the unitary gauge \cite{kaiser}. Our scope is to show that it is the multifield nature of the Higgs field (and not the gauge fields) that leads to late-time acceleration. It is important to stress that we are considering the Higgs and the gauge fields as background fields, in the sense that they affect the evolution of the scale factor via classical equations of motion. The particle content of the theory is obtained upon canonical quantization of the local fluctuations of these fields, which do not need to reflect the global spacetime symmetries. We will not discuss here the effects of the time evolution of the  background fields on the particle content of the theory.

In the next section, we lay down the general equations for the Higgs field coupled to gravity on a flat Robertson-Walker metric. In section 3 we show qualitatively how these equations describe an accelerated Universe at late time. In section 4 we solve the equations numerically and confirm the analytical predictions. Finally, we conclude in section 5 with some remarks. 

\section{Symmetries  of the Higgs model on a cosmological background}

We begin by reviewing the essential features of Higgs inflation, where the inflaton is identified with the Higgs field non-minimally coupled to gravity  \cite{hinfl}. The Lagrangian reads
\bea\label{jframe}
{{\cal L_{J}}\over \sqrt{g}}=\left({M^{2}_{p}\over 2}+\xi {\cal H}^{\dagger}{\cal H}\right)R-(D_{\mu}{\cal H})^{\dagger}(D^{\mu}{\cal H})-{1\over 4}F^{2}-V({\cal H}^{\dagger}{\cal H}),
\eea
where ${\cal H}$ is the complex Higgs doublet,  $M_{p}$ is the Planck mass, $V$ is the usual ``mexican hat'' potential $V=(\lambda/4)\left({\cal H}^{\dagger}{\cal H}-v^{2}\right)^{2}$, $D_{\mu}$ are the gauge covariant derivatives, and $F^{a}_{\mu\nu}$ is the gauge field strength. We now apply the conformal rescaling
\bea\label{conf}
g_{\mu\nu}\rightarrow \Omega^{2}g_{\mu\nu},\quad \Omega^{2}=1+{2\xi {\cal H}^{\dagger}{\cal H}\over M_{p}^{2}},
\eea
 we neglect the gauge fields, and we impose the unitary gauge so that ${\cal H}$ is replaced by the single real scalar field $h$. Then, the rescaled potential $\Omega^{-4}V$ is nearly flat in the regime $\xi h^{2}\gg M^{2}_{p}$ (which is not planckian if $\xi$ is sufficiently large) and it acts as an effective cosmological constant  for a  period of inflation sufficiently long to cope with the latest observations  \cite{planck}. The inflationary slow-roll parameters turn out to be directly related to  $\xi$  and, from their measured values, one finds that $\xi\sim 49000\sqrt{\lambda}$,  where $\lambda$ is the quartic self-interacting coupling constant of the Higgs field.

Now, let us look at the gauge fields. Since the Einstein equations are diagonal in an isotropic Universe, the off-diagonal terms of the energy momentum tensor associated to the gauge fields must vanish. The only way to achieve this is to impose the following gauge
\bea\label{ansatz}
A_{\,\,0}^{b}=0,\quad A_{\,\,i}^{b}=\delta^{b}_{\,\,i}\,f(t),
\eea
where $b$ is the gauge index while $i$ is the spatial one. The function $f(t)$ represents the only degree of freedom allowed from the gauge sector, compatible with the symmetries of the spacetime metric. With this choice,  the energy momentum tensor is diagonal by construction \cite{henneux}-\cite{Moniz:1991kx}.

We now consider the low-energy limit, when $\Omega^{2}\sim 1$ and the Jordan frame becomes indistinguishable from the Einstein frame, so the Lagrangian  \eqref{jframe}  reads
\bea\label{eframe}
{{\cal L_{E}}\over \sqrt{g}}\simeq{M_{p}^{2}\over 2}R-(D_{\mu}{\cal H})^{\dagger}(D^{\mu}{\cal H})-{1\over 4}F^{2}-V({\cal H}^{\dagger}{\cal H})\,.\eea
With the gauge \eqref{ansatz} and the  metric
\bea\label{metric}
 ds^{2}=-N^{2}(t)dt^{2}+a(t)^{2}d\vec x^{2}\, ,
 \eea
 we find that the Yang-Mills sector reduces to
\bea\label{effF}
-{1\over 4}F^{2}={3\dot f^{2}\over 2N^{2}a^{2}}-{3f^{4}\over 4a^{4}},
\eea
while the temporal gauge $A_{\,\,0}^{b}=0$ implies that $(D_{\mu}{\cal H})^{\dagger}(D^{\mu}{\cal H})\rightarrow -N^{-2}(\dot{\cal H}^{\dagger})(\dot{\cal H})$.  By variation of the fields, we derive the equations of motion in the usual way and we find that the function $f(t)$ defined in \eqref{ansatz} satisfies the simple equation  $f^{4}+2a^{2}\dot f^{2}=K_{f}=$ const. As a result, the Friedmann equations are
\bea\label{Feq}
\dot H&=&-{1\over 2 M_{p}^{2}}\left[ \dot{\cal H}^{\dagger}\dot{\cal H} +{K_{f}\over a^{4}}+\rho(1+\omega)\right],\\
H^{2}&=&{1\over 3M_{p}^{2}}\left[{1\over 2} \dot{\cal H}^{\dagger}\dot{\cal H}+ V  +{3K_{f}\over 4a^{4}}+\rho\right],
\eea 
where $H=\dot a/a$ is the Hubble parameter. We also added the contribution of an ordinary matter or radiation fluid, which satisfies the equation of state $\dot \rho+3H\rho(\omega+1)=0$. Not so surprisingly, we see that the gauge contribution is proportional to $a^{-4}$, which means that it effectively acts as radiation and, thus, cannot drive the current acceleration, nor the initial inflation. This is presumably the reason why gauge fields are neglected in Higgs inflation \footnote{Note that the contribution of the Yang-Mills term in the Friedmann equations does not change at inflationary energies, when the non-minimal coupling to gravity becomes large.}. Although the degrees of freedom of the Yang-Mills sector are completely determined by the symmetries of the background metric, the Higgs sector still enjoys an internal symmetry. We now see that this is the fundamental ingredient for accelerated expansion.

\section{Higgs dark energy}

The Higgs doublet has a $SU(2)$ (global) symmetry that we can represent by setting  \footnote{The case of $O(N)$ symmetry group was considered in \cite{zi}.}
\bea
{\cal H}={1\over \sqrt{2}}\left(\begin{array}{c}\phi_{1}\,e^{i\theta_{1}}\\  \phi_{2}\,e^{i\theta_{2}}  \end{array}\right),
\eea
where $\phi_{1,2}$ and $\theta_{1,2}$ are real functions of time. Thus, we have two Klein-Gordon equations that read
\bea\label{KG}
\ddot\phi_{i}+3H\dot\phi_{i}-{Q_{i}^{2}\over a^{6}\phi_{i}^{3}}+{\partial V\over \partial \phi_{i}}=0,\quad i=1,2,
\eea
where now 
\bea\label{potfin}
V={\lambda\over 4}(\phi_{1}^{2}+\phi_{2}^{2}-v^{2})^{2}\,,
\eea
and the constants $Q_{i}$ come from the first integrals
\bea
\dot\theta_{i}={Q_{i}\over a^{3}\phi_{i}^{2}},\quad i=1,2,
\eea
obtained by variation of the Lagrangian \eqref{eframe} with the metric \eqref{metric}. With these settings, the Friedmann equations become
\bea\label{Hsq}
H^{2}&=&{1\over 3M^{2}}\left[\frac12 \sum_{i=1,2}\left( \dot\phi_{i}^{2}+{Q_{i}^{2}\over a^{6}\phi_{i}^{2}}\right)+V+\rho_{\rm m}+\rho_{\rm r} \right]\,,\\\label{Hdot}
\dot H&=-&{1\over 2M^{2}}\left[ \sum_{i=1,2}\left( \dot\phi_{i}^{2}+{Q_{i}^{2}\over a^{6}\phi_{i}^{2}}\right)+\rho_{\rm m}+\frac43\rho_{\rm r} \right]\,,
\eea
where we have added a dust term and a radiation term that includes the Yang-Mills contribution. Respectively, these satisfy the equations
\bea
\dot \rho_{\rm r}=-4H\rho_{\rm r},\quad \dot \rho_{\rm m}=-3H\rho_{\rm m}.
\eea
The constants $Q_{i}$ are also related to the conserved $SU(2)$ current  $j^{\mu}=ig^{\mu\nu}({\cal H}^{\dagger}\nabla_{\nu}{\cal H}-\nabla_{\nu}{\cal H}^\dagger{\cal H})$, that yields in fact $Q_{1}+Q_{2}=$ const. Another quantity usually studied is the effective equation of state for the scalar fields that reads
\bea\label{omphi}
\omega_{\phi}={\dot\phi_{1}^{2}+\dot\phi_{2}^{2}+{Q_{1}^{2}\over a^{6}\phi_{1}^{2}}+{Q_{2}^{2}\over a^{6}\phi_{2}^{2}}-2V\over \dot\phi_{1}^{2}+\dot\phi_{2}^{2}+{Q_{1}^{2}\over a^{6}\phi_{1}^{2}}+{Q_{2}^{2}\over a^{6}\phi_{2}^{2}}+2V}.
\eea
As we will see below, this quantity is not very useful in the present case, as the it does not reflect the dark energy dynamics. 

The crucial role for dark energy is played by the conserved charges as it becomes apparent when one looks at the Klein-Gordon equations \eqref{KG}. As for inflationary cosmology, there are specific slow-roll conditions that can assess whether cosmic acceleration arises at late time. In the usual case of a single field and vanishing charges $Q_{1,2}$ the relevant quantities to compute are
\bea
\epsilon={M^{2}\over 2}\left(V_{,\phi}\over V\right)^{2},\quad \eta={M^{2}V_{,\phi\phi}\over V},
\eea 
where $V_{,\phi}=dV/d\phi$. Standard calculations \cite{DEbook} show that the condition  $\epsilon\ll1$ and $|\eta|\ll 1$ implies $\dot\phi^{2}\ll V$ and $|\ddot \phi|\ll|3H\dot\phi|$. In turn, this  leads to $\omega_{\phi}\sim -1+2\epsilon /3$, which means that the scalar field acts as a fluid with negative pressure that accelerate the expansion. In our case, the condition $\epsilon\ll 1$ cannot hold with the potential \eqref{potfin}, since 
\bea
\epsilon\propto {M^{2}(\phi^{2}_{1}+\phi_{2}^{2})\over V},
\eea
is a divergent quantity when the Higgs field settles in the minimum and the potential vanishes. However, this does not mean that late acceleration is not possible. In fact, thanks tho the presence of the conserved charge, there can be another regime, which we name ``ultraslow-roll condition'', characterised by  
\bea\label{ultraslow}
{Q_{i}^{2}\over a^{6}\phi_{i}^{3}}\simeq {\partial V\over \partial \phi_{i}}\,,\quad i=1,2\,
\eea
Physically, this condition implies that the kinetic terms associated to $\phi_{i}$ are much smaller that the kinetic (rotational) energy associated to $\theta_{i}$. It also means that the secular variation of $\phi_{i}$ in time is very small compared to the expansion of the Universe, and the last two terms are dominant over the first two  in the Klein-Gordon equations \eqref{KG}. This regime was already studied in \cite{spintessence} for the Abelian case (see also \cite{me}). We stress that the condition \eqref{ultraslow} makes sense only when $Q_{i}\neq 0$, so that $\sigma^{2}\equiv \phi_{1}^{2}+\phi_{2}^{2}=v^{2}$ \emph{is not} a solution of the Klein-Gordon equations. This allows the quantity $\sigma$ to be arbitrary close to $v$ and to yield an accelerating term of the appropriate magnitude, as we will shortly see.

\section{Dynamics in the ultraslow-roll regime}

In the previous section we have seen that cosmic acceleration is possible provided the condition \eqref{ultraslow} holds. We now study numerically the equations of motion and show that this regime is not only possible but also stable.  We first define the new dimensionless variables ($i=1,2$)
\bea\non
x_{i}&=&{\dot \phi_{i}\over \sqrt{6}MH}, \quad    y_{i}={Q_{i}\over \sqrt{6}MHa^{3}\phi_{i}},\quad   \lambda_{i}=-\sqrt{3\over 2}{M\over V}{dV\over d\phi_{i}}   \\\non
f_{i}&=&{\phi_{i}\over \sqrt{6}M} ,\quad z={{1\over MH}\sqrt{V\over 3}},\quad r={1\over MH}\sqrt{\rho_{\rm r}\over 3}\,.
 \eea
By differentiation with respect to the e-folding number $N=\ln a$, we find the autonomous system of equations ($i=1,2$)
\bea\label{xeq}
{dx_{i}\over dN}&=&(q-2)x_{i}+{y_{i}^{2}\over f_{i}}+z^{2}\lambda_{i}\,,\\
{dy_{i}\over dN}&=&y_{i}\left(q-2-{x_{i}\over f_{i}}\right)\,,\\\label{lambda}
{d\lambda_{i}\over dN}&=&\lambda_{i}\left({x_{i}\over f_{i}}+x_{1}\lambda_{1}+x_{2}\lambda_{2}\right)\,,\\\label{ef}
{df_{i}\over dN}&=&x_{i}\,,\\\label{zeq}
{dz\over dN}&=&(q+1-x_{1}\lambda_{1}-x_{2}\lambda_{2})z\,,\\\label{req}
{dr\over dN}&=&(q-1)r\,,
\eea 
where  $q=-1-\dot H/ H^{2}$ is the deceleration parameter, explicitly defined as 
\bea\label{dec}
q=-1+\frac32(1-z^{2})+\frac32(x_{1}^{2}+x_{2}^{2}+y_{1}^{2}+y_{2}^{2})+{r^{2}\over 2}\, .
\eea
Note that Eq.\ \eqref{Hsq} can now be written in the form $\Omega_{\rm m}+\Omega_{\rm r}+\Omega_{\rm de}=1$ where $\Omega_{\rm r}=r^{2}$  is the energy density of radiation, $\Omega_{\rm de}$ the one of dark energy and $\Omega_{\rm m}$ the one of dust, given explicitly by
\bea\label{omega}
\Omega_{\rm m}={\rho_{\rm m}\over 3M^{2}H^{2}}=1 -(x_{1}^{2}+x_{2}^{2}+y_{1}^{2}+y_{2}^{2}+z^{2}+r^{2})\, .
\eea
By inspection, one finds that the sum of the two equations \eqref{lambda}, combined with equations \eqref{ef} yield eq.\ \eqref{zeq}, so they can be eliminated from the system. The system \eqref{xeq}-\eqref{req} has fixed points at 
\bea
(x_{1},x_{2},y_{1},y_{2},r,z)=\left\{\begin{array}{c}(0,0,0,0,0,0) ,\\(0,0,0,0,0,\pm 1) ,\\(0,0,0,0,\pm 1,0),\end{array}\right.
\eea
for any value of $f_{i}$. Unfortunately, all these points are not hyperbolic so little can be said about their stability with analytical methods. However, as we will shortly see, numerical integration clearly shows  that the point characterized by $z=\pm 1$ is an attractor and the one with $r=\pm 1$ is unstable, at least in the ultraslow-roll regime. In order to explore the physically relevant solutions of this non-linear system it is convenient to consider two cases within the ultraslow-roll regime.

\subsection{Case $x_{1}=x_{2}=0$}

If we set $x_{1}=x_{2}=0$ for all $N$, so that $\dot\phi_{i}=0$, the system reduces to  ($i=1,2$)
\bea
{dy_{i}\over dN}&=&y_{i}\left(q-2\right)\,,\\
{dz\over dN}&=&(q+1)z\,,\\
{dr\over dN}&=&(q-1)r\,,
\eea 
In this case, only the phases of the Higgs doublet evolve in time, while the amplitudes are constant. There are four hyperbolic fixed points, at $(y_{1},y_{2},r,z)=$ $(0,0,0,0)$,$(\pm 1,0,0,0)$,$(0,\pm 1,0,0)$,$(0,0,\pm 1,0)$,$(0,0,0,\pm 1)$. Of these, only the one at $(y_{1},y_{2},r,z)=(0,0,0,1)$ is stable (by definition, $z>0$, so we do not consider the fixed point at $z=-1$). This fixed point corresponds to $\Omega_{\rm de}=1$, so we conclude that the dark energy-dominated solution is an attractor.

\subsection{Case $x_{i}=$ const}

\begin{figure}[ht!]
\centering
\includegraphics[scale=0.63]{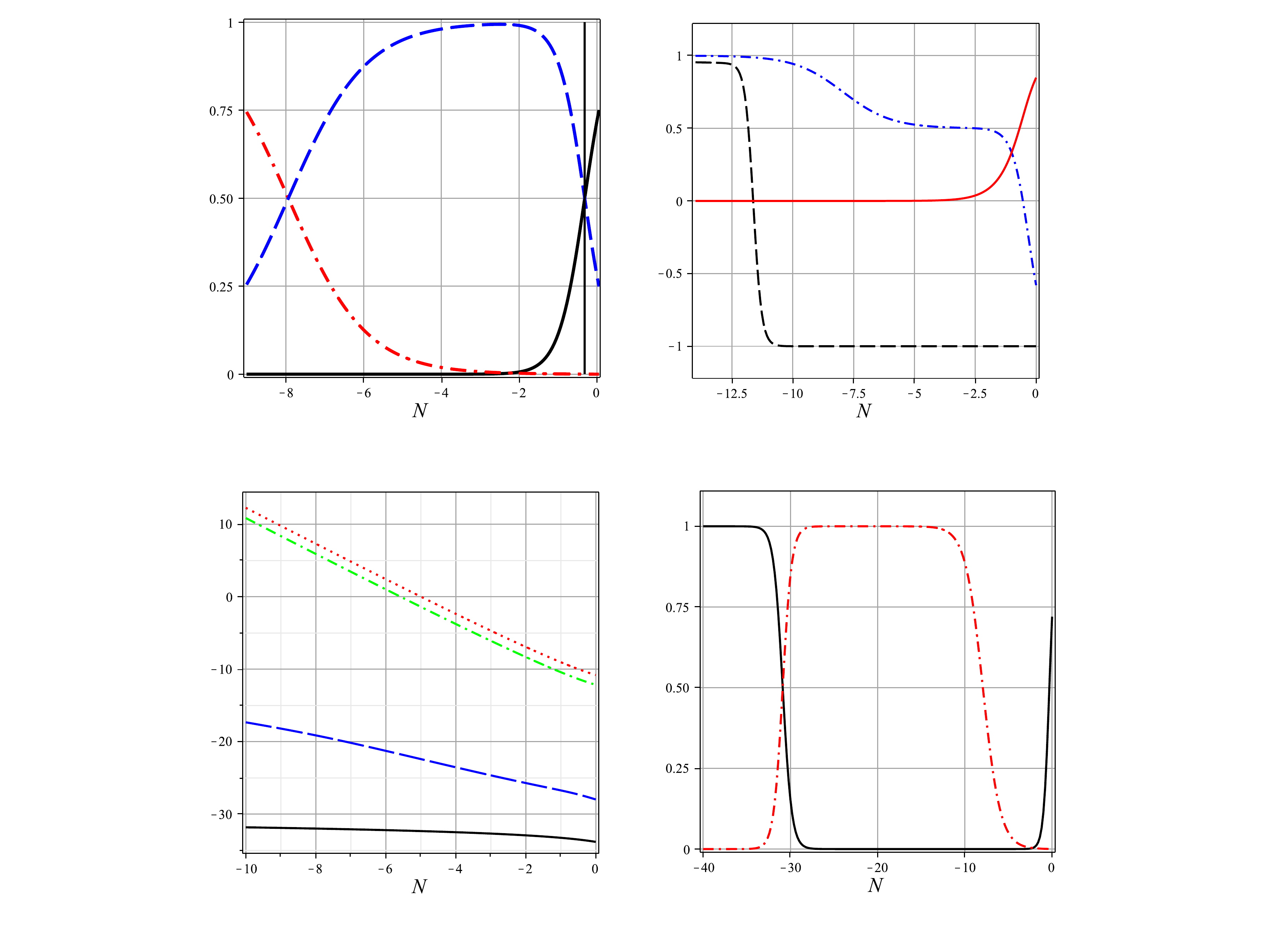}
\caption{On the left: plot of $\Omega_{\rm de}$ (solid black line), $\Omega_{\rm m}$ (blue dashed line), and $\Omega_{\rm r}$ (red dot-dashed line). The vertical grey line corresponds to $N=-0.29$ (z $\simeq -0.3$), while $N=0$ is the present time and $N\simeq -8$ (z $\simeq 3000$) is the time of matter-radiation equality. 
On the right: plot of the deceleration parameter $q$  (blue dot-dashed line) and of the effective equation of state $\omega_{\phi}$ (dashed black line). We also plot the function $z(N)$ (red solid line). The range is extended to $-14\leq N\leq 0$ to show the rise of $\omega_{\phi}$ at earlier time. }
\label{fig1}
\end{figure}

\begin{figure}[ht!]
\centering
\includegraphics[scale=0.60]{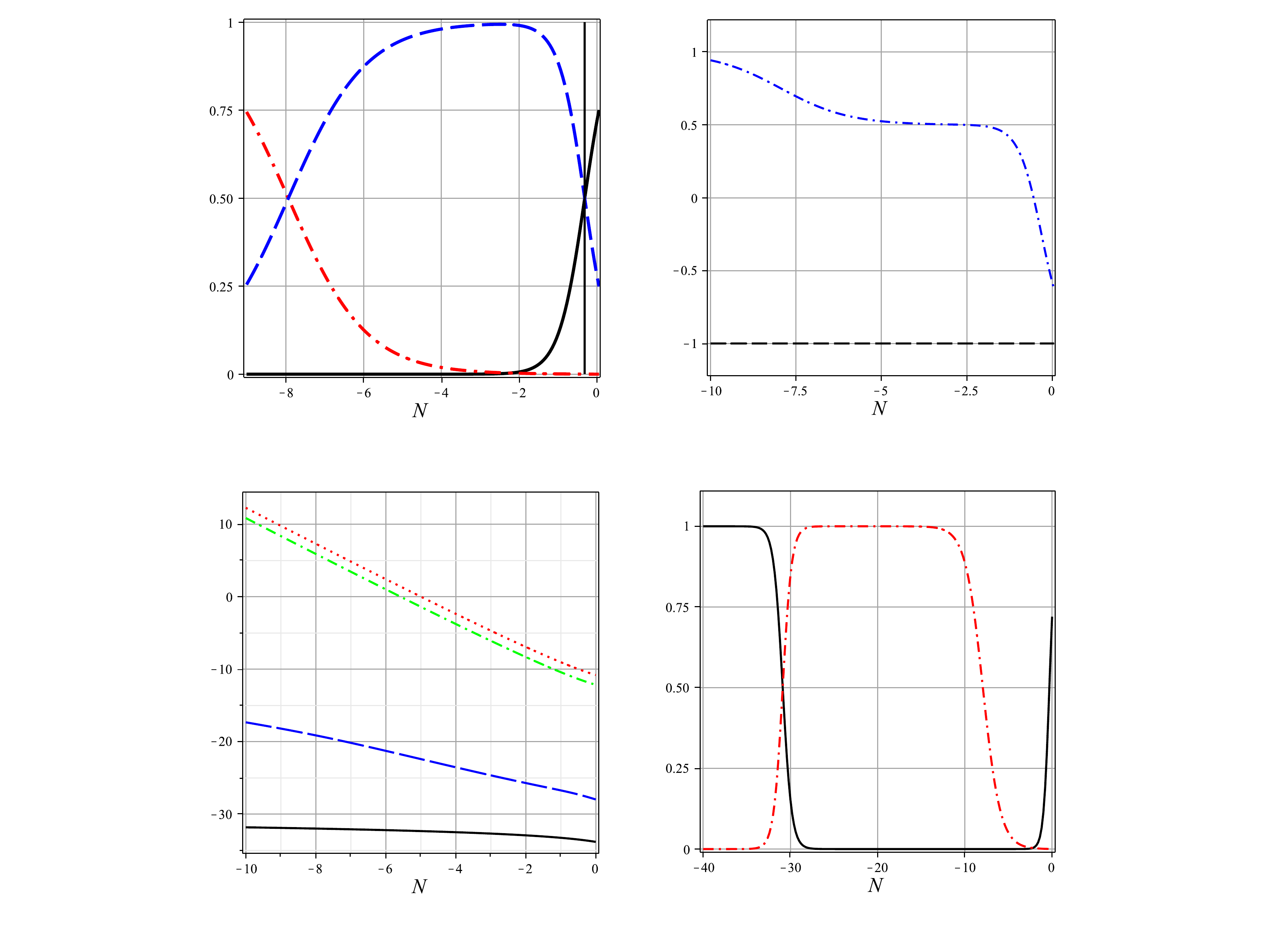}
\caption{On the left:    plot of $\log_{10} (f_{1}^{2}+f_{2}^{2})$ (black solid line), $\log_{10}(y_{1}^{2}+y_{2}^{2})$ (blue dot-dashed line),  $\log_{10}(|\lambda_{1}|)$  (red dotted line) and  $\log_{10}(|\lambda_{2}|)$  (green dash-dotted line). On the right: another plot of $\Omega_{\rm de}$ (solid black line) and $\Omega_{\rm r}$ (red dot-dashed line), numerically extrapolated for $-40<N<0$, that shows a rise of $\Omega_{\rm de}$ in the very early Universe.}
\label{fig2}
\end{figure}

A more interesting case is when we allow a non-vanishing  $\dot \phi_{i}$. To remain within the ultraslow-roll regime we impose the condition $dx_{i}/dN=0$ so that the system reduces to ($i=1,2$)
\bea
{dx_{i}\over dN}&=&0\,,\\
{dy_{i}\over dN}&=&y_{i}\left(q-2-{x_{i}\over f_{i}}\right)\,,\\
{df_{i}\over dN}&=&x_{i}\,,\\\label{eqzred}
{dz\over dN}&=&(q+1-x_{1}\lambda_{1}-x_{2}\lambda_{2})z\,,\\
{dr\over dN}&=&(q-1)r\,,
\eea 
where the functions $\lambda_{i}$ are algebraically determined by
\bea
(q-2)x_{i}+{y_{i}^{2}\over f_{i}}+z^{2}\lambda_{i}=0\,.
\eea
 The fixed points are the same as before and the one corresponding to accelerated expansion is stable. We have numerically solved the system by setting the initial conditions at the values measured today ($N=0$) of $\Omega_{\rm r}$, $\Omega_{\rm m}$, and $\Omega_{\rm de}$. We find that there is a wide range of initial conditions  that leads to the equality $\Omega_{\rm m}\simeq\Omega_{\rm de}$ at the redshift ${N}_{\rm eq}\simeq -0.29$. In figs.\ \ref{fig1} and \ref{fig2} we show the main results of the numerical computations. For these plots, we have chosen, at $N=0$, the set of initial conditions given by (we assume that $v=246$ GeV): $\phi_{1}=-v\cos (\pi/4)$, $\phi_{2}=(1+4\times 10^{-9})v\sin(\pi/4)$ so that $|\sigma-v|/v=1.6\times 10^{-9}$ and the displacement from $v$ is very small today (this choice implies that $f_{i}$ is of the order of $5\times 10^{-18}$). In order to keep small the slope of the functions $f_{i}$ we also choose $x_{1}=8\times10^{-18}$, $x_{2}=-7.5\times10^{-18}$. Finally, we set $y_{1}=8\times 10^{-10}$, $y_{2}=1.2\times 10^{-10}$, $\Omega_{\rm r}=10^{-4}$, and  $\Omega_{\rm de}=0.72$. We have observed that by changing of few order of magnitudes the initial values of $y_{i}$ only slightly affect the global evolution. On the left of  fig.\ \ref{fig1} we have plotted the densities in function of $N$ while, on the right, we have plotted the deceleration parameter $q$, the effective equation of state for the Higgs field $\omega_{\phi}$, and the function $z(N)$. We recall that the relation between $N$ and the redshift ${\rm z}$ is $N=-\ln({\rm z}+1)$ hence the domination of dark energy begins at around z $=0.34$, a  value weakly dependent on the initial conditions. By pushing the integration beyond $N=0$, we find that $\Omega_{\rm de}\rightarrow 1$ while all the other components vanish. This limit corresponds to the fixed point at $z=1$, which we know to be stable. We also find that the function $\omega_{\phi}$ does not reflect the accelerating behaviour of the Universe. In fact, this function is almost constant and equal to $-1$ for most of the evolution of the Universe. This is consistent with the fact that, in the presence of conserved charges, the global dynamics is much more complicated than in the case of a single real scalar field. We note that $\omega_{\phi}$ rapidly increases to $+1$ in the remote past, for $N<10$. However, our  solution can no longer be fully trusted beyond that point, as numerical instabilities arise. 
 
In fig.\ \ref{fig2} we plot, one the left, the functions $\log_{10} (f_{1}^{2}+f_{2}^{2})$, $\log_{10} (y_{1}^{2}+y_{2}^{2})$,  and  $\log_{10} (|\lambda_{i}|)$. On the right, we have pushed the numerical integration beyond matter-radiation equality with the surprising result that the energy density of dark energy dominates before some value of $N$ that depends on the initial conditions. However, as before, we cannot fully trust our solutions in this region and more powerful numerical techniques are required.
 
\section{Conclusions}

In summary, we have shown that the Higgs model coupled  to gravity not only leads to a successful inflationary model, but it can also explain the current accelerated expansion. The fundamental mechanism is the $SU(2)$ symmetry of the background Higgs field that carries an associated charge. In turn, the charge enter the Friedmann equations, causing accelerated expansion, similarly to the ``spintessence'' model. The magnitude of the acceleration is related to the displacement of the Higgs field from its vacuum, which, together with the charges, act as an effective and slowly varying cosmological constant, vanishing in the infinite future. In short, by chasing its own vacuum, the Higgs field accelerates the Universe. 

We have confirmed our findings with the numerical analysis of the equations of motion that form a non-linear system. The latter is rather complicated with several non-hyperbolic fixed points and classes of solutions. Among these, one implies a stationary Higgs field squared ${\cal H}^{\dagger}{\cal H}$ and, provided the associated charges are non-vanishing, the late accelerating Universe is an attractor solution. Another class of solutions is given by linearly (in $N$) evolving fields $\phi_{i}$. This is a more realistic solution as it implies a secular variation of the background Higgs field. Our numerical analysis shows that this solution is well-behaved and robust against variations of initial conditions. The most general numerical solution of the system is much more complicated and it is postponed to future work. However, we believe that the classes of solutions examined so far show that the non-Abelian Higgs field can be a viable  source of dark energy.

These considerations are purely classical, in the sense that they do not concern the physical particle content of the theory, which is determined exclusively by the canonical quantization of the fluctuations of the fields around a classical solution of eqs.\ \eqref{xeq}-\eqref{req}. This is a highly non-trivial task when the space-time curvature is not negligible. In fact, during inflation, the fluctuations of the Goldstone bosons mix with the ones of the metric and loop corrections are modified, see e.g. \cite{postma,burgess}. At lower energy, however, field perturbations decouple from gravity so they can be quantized canonically as in flat space. In particular, the standard electroweak symmetry breaking occurs as usual. This does not imply, however, that the  Goldstone bosons disappear from the classical theory. For example, in the Abelian case, the correct interpretation of symmetry breaking is that the time-like component of the massless photon is canceled by the scalar field only at the quantum level (in analogy with the Gupta-Bleurer mechanism). This is particularly transparent in the so-called $R_{\xi}$ gauge formalism, see e.g. Chapter 21 of \cite{peskin}.  At the classical level, the Goldstone bosons still satisfies the Klein-Gordon equations \eqref{KG} and this is all it matters for the evolution of the background fields and, in particular, of the metric.

We believe that this model is very promising since it does not require any exotic form of dark energy. It is just the background Higgs that can also lead inflation. Of course, to verify that this is a true and viable model of dark energy, we need a further step, namely a detailed study of the the classical perturbations together with the analysis of the most general solution to the equations of motion, which will be the focus of future work.

I would like to thank A.\ F\"uzfa, S.\ Zerbini, L.\ Vanzo, and G.\ Cognola for fruitful discussions and valuable comments. 
 
{}


\section*{References}
\begin{thebibliography}{99}


\bibitem{inflation}
A.~D.~Linde,
  Phys.\ Lett.\ B {\bf 129} (1983) 177;
A.~H.~Guth,
  Phys.\ Rev.\ D {\bf 23} (1981) 347;
  V.~F.~Mukhanov and G.~V.~Chibisov,
  JETP Lett.\  {\bf 33} (1981) 532;
 A.~A.~Starobinsky,
  Phys.\ Lett.\ B {\bf 91} (1980) 99.

 \bibitem{planck}
 P.~A.~R.~Ade {\it et al.}  [Planck Collaboration],
  arXiv:1303.5082 [astro-ph.CO].
  
  \bibitem{euclid}
L.~Amendola {\it et al.}  [Euclid Theory Working Group Collaboration],
  Living Rev.\ Rel.\  {\bf 16} (2013) 6
  
  
  
    \bibitem{hinfl}
  F.~L.~Bezrukov and M.~Shaposhnikov,
  Phys.\ Lett.\ B {\bf 659} (2008) 703;
  R.~Fakir and W.~G.~Unruh,
  Phys.\ Rev.\ D {\bf 41} (1990) 1783.
  


\bibitem{monop}
 S.~Schlogel, M.~Rinaldi, F.~Staelens, and A.~Fuzfa,
  Phys.\ Rev.\ D {\bf 90} (2014) 044056;
A.~Fuzfa, M.~Rinaldi, and S.~Schlogel,
  Phys.\ Rev.\ Lett.\  {\bf 111} (2013) 12,  121103.
  
  
    
  \bibitem{achu}
  A.~Achucarro and T.~Vachaspati,
  Phys.\ Rept.\  {\bf 327} (2000) 347.
  
  

  
  
  \bibitem{henneux}
   M.~Henneaux,
  J.\ Math.\ Phys.\  {\bf 23} (1982) 830.

  
  

 
\bibitem{picon}
C.~Armendariz-Picon,
  JCAP {\bf 0407} (2004) 007.

\bibitem{Cembranos:2012ng}
  J.~A.~R.~Cembranos, A.~L.~Maroto and S.~J.~Nœ–e.~Jare–o,
  Phys.\ Rev.\ D {\bf 87} (2013) 4,  043523.
  
   \bibitem{galtsov}
  D.~V.~Galtsov and M.~S.~Volkov,
  Phys.\ Lett.\ B {\bf 256} (1991) 17.
  
  \bibitem{Moniz:1991kx}
  P.~V.~Moniz, J.~M.~Mourao and P.~M.~Sa,
  Class.\ Quant.\ Grav.\  {\bf 10} (1993) 517.
  
  \bibitem{Maleknejad:2012fw}
  A.~Maleknejad, M.~M.~Sheikh-Jabbari and J.~Soda,
  Phys.\ Rept.\  {\bf 528} (2013) 161.
  
  \bibitem{Elizalde:2003ku}
  E.~Elizalde, J.~E.~Lidsey, S.~Nojiri and S.~D.~Odintsov,
  Phys.\ Lett.\ B {\bf 574} (2003) 1.
  
  \bibitem{Adshead:2012kp}
  P.~Adshead and M.~Wyman,
  Phys.\ Rev.\ Lett.\  {\bf 108} (2012) 261302.
  
  \bibitem{Fuzfa}
  A.~Fuzfa and J.~-M.~Alimi,
  Phys.\ Rev.\ D {\bf 73} (2006) 023520.
  
  
  
  
  

  
\bibitem{kaiser}
R.~N.~Greenwood, D.~I.~Kaiser and E.~I.~Sfakianakis,
  Phys.\ Rev.\ D {\bf 87} (2013) 6,  064021;
K.~Schutz, E.~I.~Sfakianakis and D.~I.~Kaiser,
  Phys.\ Rev.\ D {\bf 89} (2014) 064044;
  D.~I.~Kaiser and E.~I.~Sfakianakis,
  Phys.\ Rev.\ Lett.\  {\bf 112} (2014) 011302.
  
   
 
  \bibitem{zi}  
X.~-z.~Li, J.~-g.~Hao and D.~-j.~Liu,
  Class.\ Quant.\ Grav.\  {\bf 19} (2002) 6049.
  
  \bibitem{DEbook}
  L.\ Amendola and S.\ Tsujikawa, ``Dark energy - theory and observations", Cambridge University Press, 2010.
  
   \bibitem{spintessence}
  L.~A.~Boyle, R.~R.~Caldwell and M.~Kamionkowski,
  Phys.\ Lett.\ B {\bf 545} (2002) 17.
  
   \bibitem{me}
  M.~Rinaldi,
  Eur.\ Phys.\ J.\ Plus 129 (2014)  56.
  
     
 
 \bibitem{postma}
   S.~Mooij and M.~Postma,
  JCAP {\bf 1109} (2011) 006.
  
  \bibitem{burgess}
C.~P.~Burgess, H.~M.~Lee and M.~Trott,
  JHEP {\bf 1007} (2010) 007.
  
  \bibitem{peskin}
  M.~E.~Peskin and D.~V.~Schroeder,
  ``An Introduction to quantum field theory,''
  Reading, USA: Addison-Wesley (1995) 842 p.


\end{thebibliography}
\end{document}